# Was there COVID-19 back in 2012? – Challenge for AI in Diagnosis with Similar Indications

Imon Banerjee, PhD[1,5], Priyanshu Sinha[2], Saptarshi Purkayastha, PhD[3], NazaninMashhaditafreshi, BSc[4], Amara Tariq, PhD[1], Jiwoong Jeong, MS[1], Hari Trivedi, MD[1,5], Judy W. Gichoya, MBChB MS[1,5]

[1]Department of Biomedical Informatics, Emory School of Medicine, Atlanta, USA;
[2]MentorGraphics Indian Pvt. Ltd., India;
[3]School of Informatics & Computing, Indiana University, Purdue University, Indianapolis, USA;
[4]Department of Computer Engineering, K. N. Toosi University of Technology, Tehran, Iran;
[5]Department of Radiology, Emory School of Medicine, Atlanta, USA

*Abstract*

Purpose: Since the recent COVID-19 outbreak, there has been an avalanche of research papers applying deep learning based image processing to chest radiographs for detection of the disease. To test the performance of the two top models for CXR COVID-19 diagnosis on external datasets to assess model generalizability.

Methods: In this paper, we present our argument regarding the efficiency and applicability of existing deep learning models for COVID-19 diagnosis. We provide results from two popular models - COVID-Net and CoroNet evaluated on three publicly available datasets and an additional institutional dataset collected from EMORY Hospital between January and May 2020, containing patients tested for COVID-19 infection using RT-PCR.

Results: There is a large false positive rate (FPR) for COVID-Net on both ChexPert (55.3%) and MIMIC-CXR (23.4%) dataset. On the EMORY Dataset, COVID-Net has 61.4% sensitivity, 0.54 F1-score and 0.49 precision value. The FPR of the CoroNet model is significantly lower across all the datasets as compared to COVID-Net - EMORY(9.1%), ChexPert (1.3%), ChestX-ray14 (0.02%), MIMIC-CXR (0.06%).

Conclusion: The models reported good to excellent performance on their internal datasets, however we observed from our testing that their performance dramatically worsened on external data. This is likely from several causes including overfitting models due to lack of appropriate control patients and ground truth labels. The fourth institutional dataset was labeled using RT-PCR, which could be positive without radiographic findings and vice versa. Therefore, a fusion model of both clinical and radiographic data may have better performance and generalization.

**Abbreviations**

DL = deep learning, CNN = convolutional neural network, COVID-19 = Coronavirus disease 2019

**Summary**

Existing COVID-19 detection models for chest radiographs demonstrate extremely poor performance on datasets that pre-dated the discovery of COVID-19. They also perform poorly on a new institutional dataset that contains confirmed COVID-19 cases.

**Key Points**

- Two existing models for detection of COVID-19 on chest radiographs were tested on three public datasets and one institutional dataset.

- There is an extremely high rate of false positive detection of COVID-19 (up to 61%) for one model on datasets created before COVID-19.

- There is a very high false negative rate (94%) for a second model tested on confirmed COVID-19 cases from our institution.

- This may be due to a combination of factors, including overfitting of models on training datasets, similar appearance of COVID-19 to other infections, and a portion of serology positive COVID-19 cases that do not demonstrate any chest radiograph findings.

**Introduction**

As of May 2020, the number of global confirmed cases of COVID-19 was over 3.5 million spanning 187 territories and six continents. The United States reported more than 1 million cases, with more than 80,000 deaths over a span of 3 months. Given the lack of adequate testing kits, curative treatment, or availability of a vaccine, mitigation strategies were implemented both in hospitals such as rescheduling of elective cases and limiting visitation; and outside of hospitals in the form of contact tracing and social distancing in order to "flatten the curve." The gold standard for SARS-CoV-2 testing is Reverse Transcription Polymerase Chain Reaction (RT-PCR) [1] with a turnaround time of 4 to 48 hours. Due to the unavailability of rapid testing, both computed tomography (CT) and chest radiographs (CXR) have been utilized in lieu of or as a stopgap for laboratory testing. Even though routine screening of patients with imaging is not recommended [2, 3, 4], these exams are still frequently obtained as a test for disease in the outpatient and emergency department (ED) settings, and therefore there is value in leveraging any technology to expedite review or improve the quality of reads for these cases.

Deep learning (DL) based architectures have been applied to medical imaging with tremendous success, including classification and object localization on CXR, with reported expert-level performance for detection of various chest pathologies [5, 6]. With respect to COVID-19, much work has been done recently to detect findings of the disease on chest radiographs as a means to rapidly screen a high number of patients[7]. However a major gap in this work is that the clinical utility of these models has not been tested on large, external datasets. DL models trained on narrow data are known to underperform on external data for various reasons including differences in image acquisition protocols[8, 9], image processing[10], distribution pipelines (e.g., image compression[11]), and overlooked pathology[12]. For AI to be clinically useful, models must generalize to unseen data from different populations. Otherwise there is potential for models to perpetuate already existing biases in their training datasets, even exacerbating racial disparities already seen in COVID-19.

To assess clinical utility of AI applied to CXRs for COVID-19, we selected two recently developed open-source deep learning based COVID-19 detection models - COVID-Net[13] and CoroNet[14], and applied these models to four CXR datasets. Three of these datasets were collected before the spread of COVID-19 and the fourth institutional dataset was collected between Jan - May, 2020. We present our in-depth analysis using statistical reasoning and qualitative assessment by trained radiologists to assess the results of these existing models on external datasets.

**Materials and Methods**

*Chest radiograph Datasets*

In this section, we present details of the CXR datasets used in our experiments. Table 1 summarizes the basic demographics of the population in the four datasets and the sections below describe the datasets. Ground truth label distribution of the datasets are shown as bar charts in Figure 1.

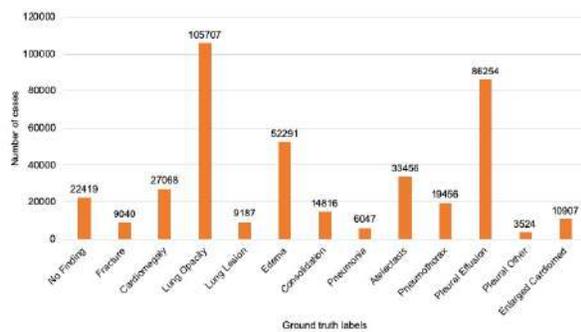

(a)

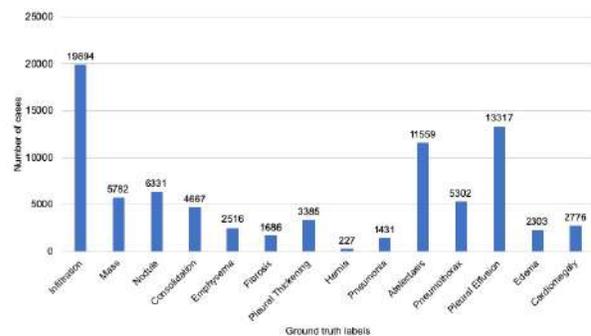

(b)

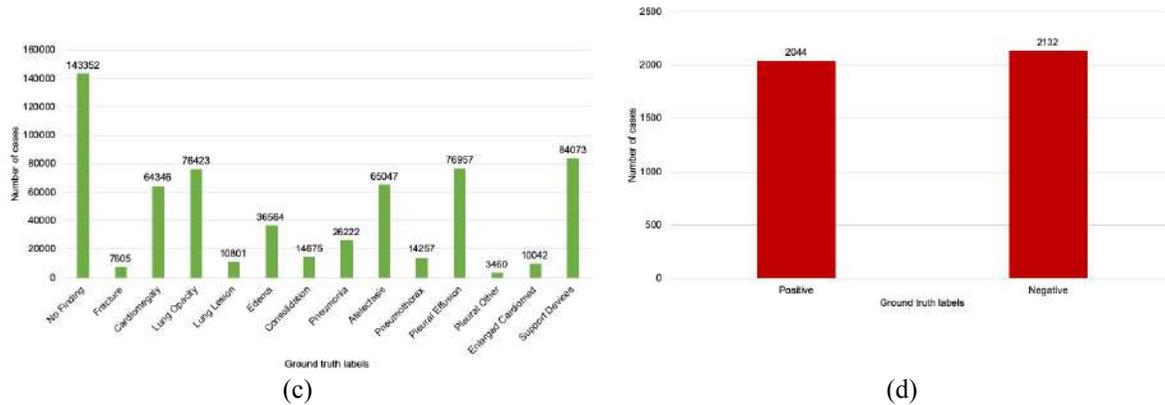

(c)                          (d)

**Figure 1:** Distributions of the ground truth labels in datasets used in the study: (a) Label distribution in ChexPert v1.0 Dataset; (b) Label distribution in ChestX-ray14 Dataset; (c) Label distribution in MIMIC-CXR Dataset label indicates a CXR without positive RT-PCR within 3 days or no positive RT-PCR at any time; (d) Label distribution in EMORY Dataset - A positive label indicates a positive RT-PCR within 3 days of the CXR. A negative

**Table 1.** Statistical makeup of the datasets including basic demographics information. Missing information is documented as `Not Available`. *Other pulmonary disease* refers to effusion, pleural thickening, edema etc.

|  | **EMORY Dataset** | **ChexPert** | **ChestX-ray14** | **MIMIC-CXR** |
|---|---|---|---|---|
| *Total COVID-19+ Cases* | 1,786 | 0 | 0 | 0 |
| *Age (mean/std)* | 55.9/20.0 | 60.31/18.56 | 46.13/ 17.01 | Not Available |
| *Gender* | | | | |
| *Male* | 930 | 35917 | 16630 | Not Available |
| *Female* | 753 | 28822 | 14175 | |
| *Unknown/unreported* | 103 | 1 | 0 | |
| *Race* | | | | |
| *Caucasian* | 328 | Not Available | Not Available | Not Available |
| *Black* | 821 | | | |
| *Asian* | 230 | | | |
| *Other* | 63 | | | |
| *Unknown* | 344 | | | |
| *Comorbidities* | | | | |
| *Pneumonia* | 665 | 6047 | 1413 | 26222 |
| *Other Pulmonary diseases* | 78 | 89778 | 13702 | 80417 |

Dataset#1: ChexPert v1.0 ChexPert is a large dataset containing 224316 CXRs of 65240 patients acquired from October 2002 to July 2017. The data is labeled with 14 different labels. Label distribution is shown in Figure 1a.

Dataset#2: ChestX-ray14 (previously ChestX-ray8) The ChestX-ray14 dataset includes 108948 frontal-view CXRs of 32717 unique patients acquired from 1992 to 2015. 14 disease labels were identified from the associated radiology reports using text mining. CXRs with "No Finding" are not included in the 14 labels. In this dataset, an image can have multiple labels. Label distribution is shown in Figure 1b.

Dataset#3: MIMIC-CXR The MIMIC-CXR-JPG v2.0.0 dataset contains 377110 chest radiographs associated with 227827 CXRs sourced from the Beth Israel Deaconess Medical Center between 2011 - 2016.

labels extracted from associated free-text radiology reports using natural language processing. Label distribution is shown in Figure 1c.

Dataset#4: EMORY Dataset (Institution names hided for blind review) With the approval of EMORY Institutional Review Board, we extracted 4176 CXRs from 1,786 patients treated across four EMORY Healthcare hospitals between January - May 2020 and de-identified the scans. For this dataset, we consider an exam as COVID-19 positive if real-time reverse transcriptase-polymerase chain reaction (RT-PCR) results were positive within a 3 day window of the CXR exam date. A positive label corresponds to the RT-PCR test result, and does not necessarily mean that the radiograph has imaging findings of COVID-19.

*Models*

In this section, we summarize two recently published models - COVID-Net [13] and CoroNet [14] used in our evaluation and provide details of our experimentation methodology.

COVID-Net [13] is a convolutional neural network architecture with a long range of connectivity, designed by researchers from University of Waterloo, Canada. This network is specifically designed to maximize the sensitivity to COVID-19 to limit the number of missed cases. The architecture includes some innovations. First, it is a light-weight network with 117,437,315 trainable parameters which follows a residual projection-expansion- projection-extension (PEPX) design pattern. In PEPX, a 1×1 convolution projects features into a lower dimension followed by a second 1×1 convolution that expands the features into a higher dimension. Second, the architecture sparingly leverages long range connections which act as central hubs for earlier layers to connect to much later layers in the network. The model was trained on the COVIDx dataset which is comprised of a total of 13,975 CXR images from 13,870 patient cases Table 1: Statistical makeup of the datasets including basic demographics information. Missing information is documented as *'Not Available. Other pulmonary disease* refers to effusion, pleural thickening, edema etc. with learning rate=2e-4, number of epochs=22, batch size=64, factor=0.7, and patience=5. The COVIDx dataset [15] is a collection of five different open-source datasets - 1) COVID-19 Image Data Collection, 2) COVID-19 Chest X-ray Dataset Initiative, 3) ActualMed COVID-19 Chest X-ray Dataset, 4) RSNA Pneumonia Detection Challenge dataset, and 5) COVID-19 radiography database. The authors used traditional data augmentation for training: translation, rotation, horizontal flip, zoom, and intensity shift. Reported performance was 93% accuracy and 91% sensitivity.

CoroNet [14] is a two stage deep learning model which was developed by researchers at the University of Illinois at Chicago. The first module called a *feature extractor* uses AutoEncoders to extract the relevant features and information from the raw data. The next module performs image classification based on convolution operations on the extracted residual features from the feature extractor module. The classification module uses a ResNet-18 architecture pretrained on ImageNet. The model was also trained on the COVIDx dataset and the authors used weighted entropy loss to address class imbalance. With this two-stage architecture with unsupervised feature extraction, they generated a total of 11.8M trainable parameters which is 10 times fewer parameters than COVID-Net. The authors reported 93.5% overall class average accuracy with 93.63% precision.

*Experimentation methodology*

For inference on our four datasets, both models were executed with their original weights as provided by the authors. In the default configuration, COVID-Net is able to predict three different labels, i.e., 'COVID-19', 'Pneumonia', and 'Normal' whereas CoroNet predicts 'COVID-19' and 'Healthy'. We assume that no COVID-19 cases were present in the ChexPert (images acquired from October 2002 to July 2017), ChestX-ray14 (images acquired from 1992 to 2015), and MIMIC-CXR (images acquired from 2011 to 2016) datasets since the SARS-CoV-2 virus was discovered in December, 2019 [16]. Thus, for these datasets, we considered any case predicted as COVID-19 positive as a false positive and considered this false positive rate as a key measurement of model performance. For the EMORY dataset, we have ground truth COVID-19 labels based on RT-PCR test results so we are able to calculate both false positive and false negative prediction rates. We also provide a confusion matrix to further illustrate the model results as compared to the ground truth.

Lastly, we produced visual explanations for the results of both models on select cases using the Gradient-weighted Class Activation Mapping (Grad-CAM) algorithm [17] to understand the reasoning behind the positive COVID-19 predictions. These images were randomly sampled and shown to two board certified radiologists to understand the possible explanation of COVID-19 prediction.

All work was done on an 8x Quadro RTX 6000 CUDA enabled GPU machine with 512GB RAM.

**Results**

Table 2 presents the performance analysis of the COVID-Net and CoroNet models over the four different datasets. There is a large false positive rate (FPR) for COVID-Net on both ChexPert (55.3%) and MIMIC-CXR (23.4%) datasets, highlighting the lack of specificity in these models. Interestingly, the FPR was significantly lower for CXR14 dataset (only 0.05%) which may be explained by the presence of more cases with nodules and masses compared to ChexPert and MIMIC-CXR. Moreover, there are fewer cases marked as *lung opacity* which can be easily mistaken for COVID-19 infection (as well as other types of pneumonia (Figure 1). For EMORY dataset, COVID-Net correctly predicted 1255 COVID19 cases as positive and resulted 61.4% sensitivity. However, 1296 cases were wrongly predicted as positives and the model also missed 789 COVID-19 cases. Thus the COVID-Net model scored only 0.54 F1-score and 0.49 precision value on EMORY dataset.

**Table 2:** Distribution of the predicted labels from the COVID-Net and CoroNet models across all datasets along with validation measurements. FN: false negative, FP: false positive

| Model | Labels | EMORY dataset | ChexPert | ChestX-ray14 | MIMIC-CXR |
|---|---|---|---|---|---|
| *COVID-Net* | COVID-19<br>Pneumonia<br>Normal | 2551<br>1298<br>327 | 123610<br>90007<br>10031 | 523<br>40638<br>70959 | 88068<br>183583<br>105459 |
| | Validation | FP-rate:61.1%<br>FN-rate:38.6% | FP-rate: 55.3% | FP-rate: 0.5% | FP-rate: 23.4% |
| *CoroNet* | COVID-19<br>Healthy | 322<br>3854 | 2970<br>220678 | 21<br>112099 | 2183<br>3749279 |
| | Validation | FP-rate:9.1%<br>FN-rate:93.7% | FP-rate: 1.3% | FP-rate: 0.02% | FP-rate: 0.06% |

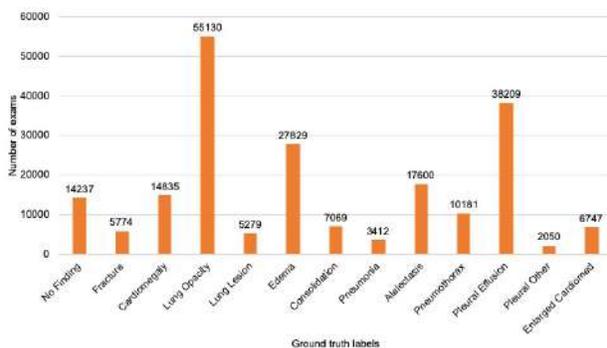 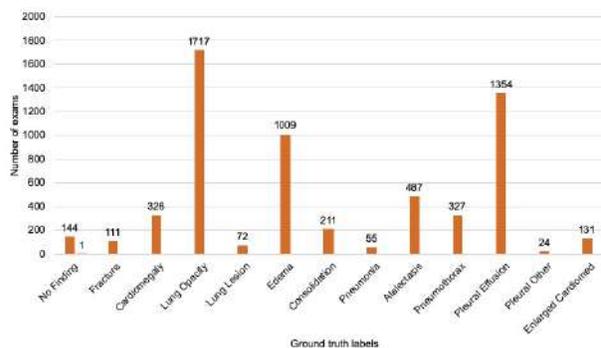

(a) COVID-Net prediction on the ChexPert v1.0 Dataset    (b) CoroNet prediction on the ChexPert v1.0 Dataset

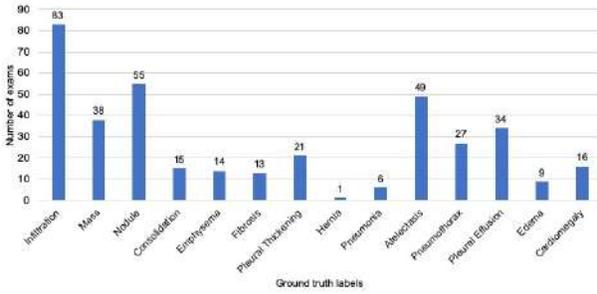
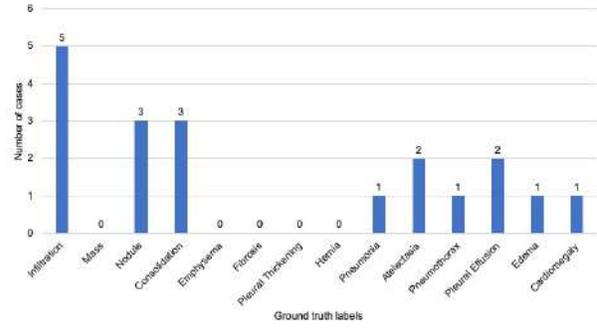

(c) COVID-Net prediction on the ChestX-ray14 Dataset    (d) CoroNet prediction on the ChestX-ray14 Dataset

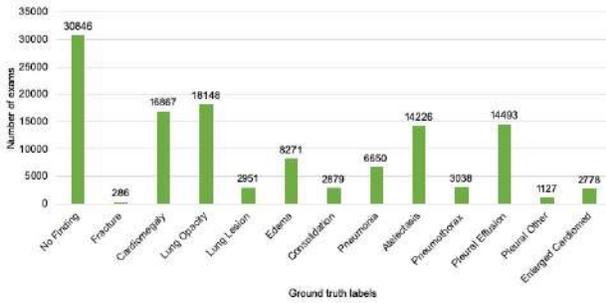
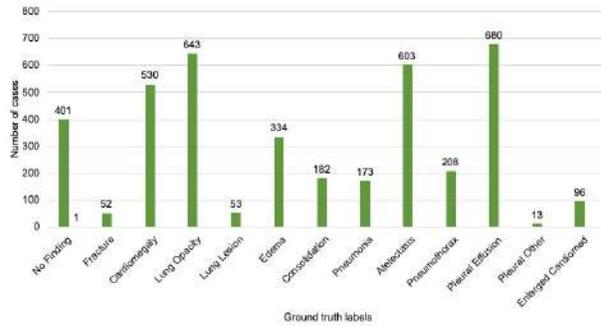

(e) COVID-Net prediction on the MIMIC-CXR dataset    (f) CoroNet prediction on the MIMIC-CXR Dataset

**Figure 2:** Distribution of positive COVID-19 predictions from the COVID-Net and CoroNet models as compared to ground truth labels from the ChexPert, ChestX-ray14, and MIMIC-CXR datasets. In the ChexPert and ChestX-ray8 datasets, a majority of FP predictions were in cases with lung opacities, infiltrations, consolidations, and effusions which are seen in a variety of infectious processes, including COVID-19.

The FPR of the CoroNet model is significantly lower across all the datasets as compared to COVID-Net - EMORY (9.1%), ChexPert (1.3%), ChestX-ray14 (0.02%), MIMIC-CXR (0.06%). Following the similar trend as of COVIDNet, only 19 cases are predicted as COVID-19 positive by CoroNet on the ChestX-ray14 dataset. However, false negative rate (FNR) on the EMORY patients dataset is extremely high 93.7% since it correctly identified on 129 COVID19 positive cases and missed 1915 cases. Table 3 presents the confusion matrix for the COVID-Net and CoroNet models on the EMORY Dataset, where RT-PCR test results (within 3 days of the radiograph) are considered as ground truth labels. This again highlights the high FPR for COVID-Net and a high FNR of CoroNet.

In order to understand the models' bias towards positive COVID-19 prediction, we plotted the distribution of 'COVID19' predicted labels against the ground truth labels of the ChexPert, ChestX-ray14, and MIMIC-CXR dataset. As seen from Figure 2, the highest number of false positive predictions by both COVID-Net and CoroNet in ChexPert cases contains lung opacity and the second highest is pleural effusion. CoroNet also predicted mostly lung opacity cases as COVID-19 positive in the MIMIC-CXR dataset. Interestingly on the same dataset, COVID-Net mostly predicted cases with no chest disease as COVID-19 positive.

**Table 3:** Confusion matrix demonstrating predictions of the COVID-Net and CoroNet on the EMORY dataset. The COVID-19 dataset demonstrates very high false positive and false negative rates, with close to even distributions for both predictions. The CoroNet dataset demonstrates a high propensity for negative predictions with a 94% false negative rate.

|  | COVID-Net | CoroNet |

| T | Labels | COVID19+ | COVID19- | COVID19+ | COVID19- |
| R | COVID19+ | 1255 | 789 | 129 | 1915 |
| U | COVID19- | 1296 | 836 | 193 | 1939 |
| T | | \multicolumn{4}{c}{P R E D I C T E D} | | | |

**Table 4:** Percentage of images with support devices detected as COVID19+ by COVID-NET

| Dataset | COVID19+ Detections in Images with support devices |
|---|---|
| *ChexPert* | 52.7% |
| *MIMIC-CXR* | 21.2% |

The Grad-CAM results help to understand the underlying reasons for a high FP rate for COVID-Net and high FN rate for CoroNet. Figure 3, 4, and 5, demonstrate that the COVID-Net model mainly focuses on the presence of support devices rather than anatomy for predicting positive COVID-19 labels, predisposing it to assign positive labels to ICU patients. Table 4 demonstrates the presence % of support devices present positive predictions for the COVID-19 positive by COVID-Net on all the datasets. Figure 6 presents the GradCAM for false negative EMORY cases (with and without support devices) for CoroNet. Even though the CoroNet model does not have an explicit bias towards support devices, in most cases the model's attention is out of focus.

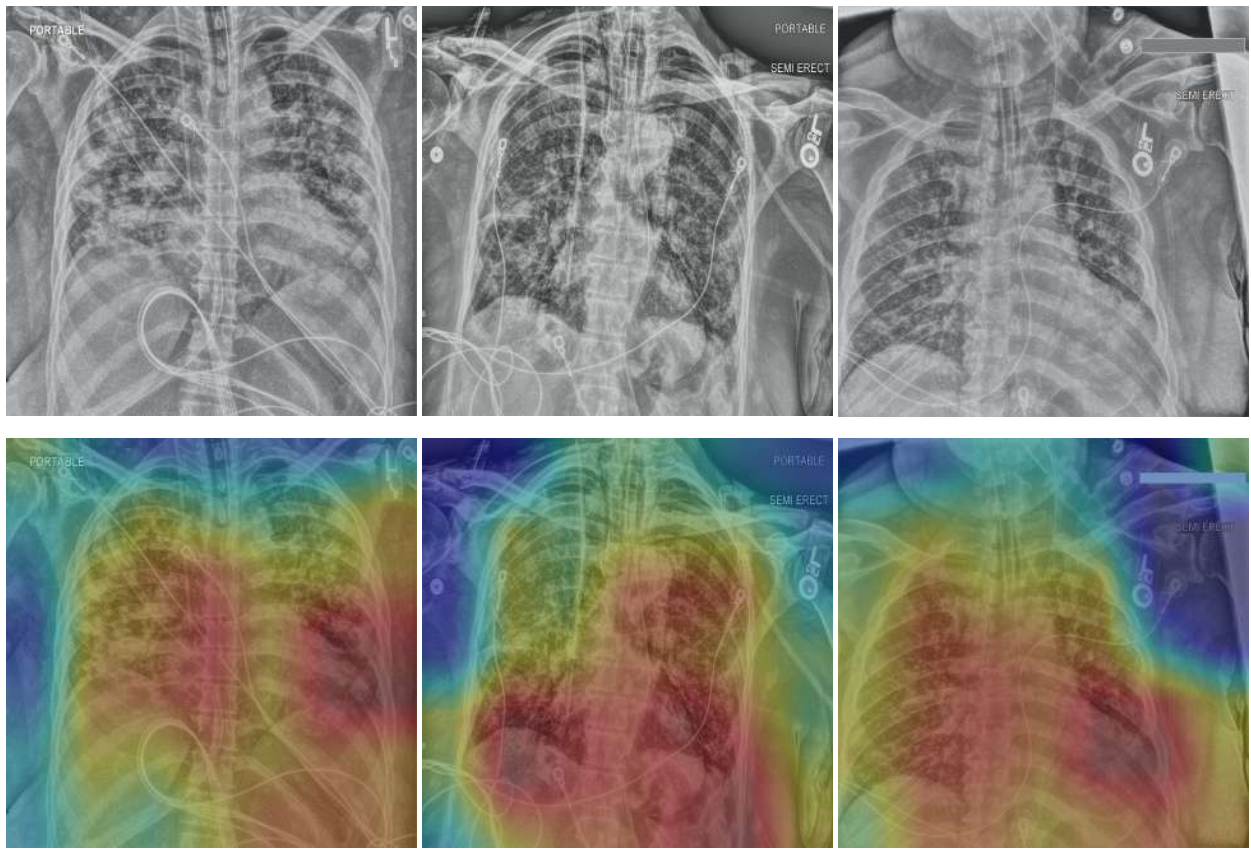

**Figure 3:** GRAD-CAM images of false positive cases detected by the COVID-net model on the EMORY dataset. All three of these radiographs contained an endotracheal tube and parenchymal findings which highlights the possible bias of the model to these characteristics

**Discussion**

For AI to be clinically useful, models must generalize to unseen data from different populations which requires adequately sized datasets collected from external sites. The COVID-Net and CoroNet models reported good to excellent performance (COVID-Net - 93 % accuracy and 91 % sensitivity; CoroNet - 93.5% accuracy and 93.63 % precision) on their internal datasets, however we observed from our testing that their performance dramatically worsened on external data. The concept of decreased performance for DL models on external datasets is a known issue and was previously described in chest radiography[18] [19] in the range of 3.6 – 5.2% performance decrease in detecting pneumonia across institutions. This relatively small drop in performance was attributed to differences in *differences in labeling*[19] as well as *data distribution* between the datasets. However, in our case the performance decreased precipitously which indicates a broader underlying issue.

The prediction of a COVID-19 diagnosis by both models before the spread of SARS-CoV2 (often by many years) highlights that these models are highly overfitted to their training and validation datasets. This issue is likely multifactorial, but we believe that primary issue is not model architecture, but rather the lack of appropriate control patients in these datasets which come mainly from the ICU rather than the general outpatient or emergency department population. Therefore the ability of these models to generalize on four broader patient datasets is severely limited, as demonstrated in our results. COVID-Net had an alarmingly high false-positive rate (up to 61 %) when tested on CXR datasets collected before the spread of virus. CoroNet had a lower false-positive rate (only 1.3 %) but instead failed to detect most confirmed COVID-19 cases in the EMORY dataset. We investigated the results of both models as compared to ground truth labels to identify the source of errors and found that instead of learning distinctive imaging features for COVID-19 diagnosis, the models mainly focused on either co-association of other findings (e.g. lung opacity, pleural effusion) or the presence of support devices. The radiographic appearance of COVID-19 is non-specific, typically described as symmetric, peripheral-predominant nodular and groundglass opacities. However this appearance is also seen in many other atypical and opportunistic infections. It is likely that the true cause of errors is somewhere in between these two reasons.

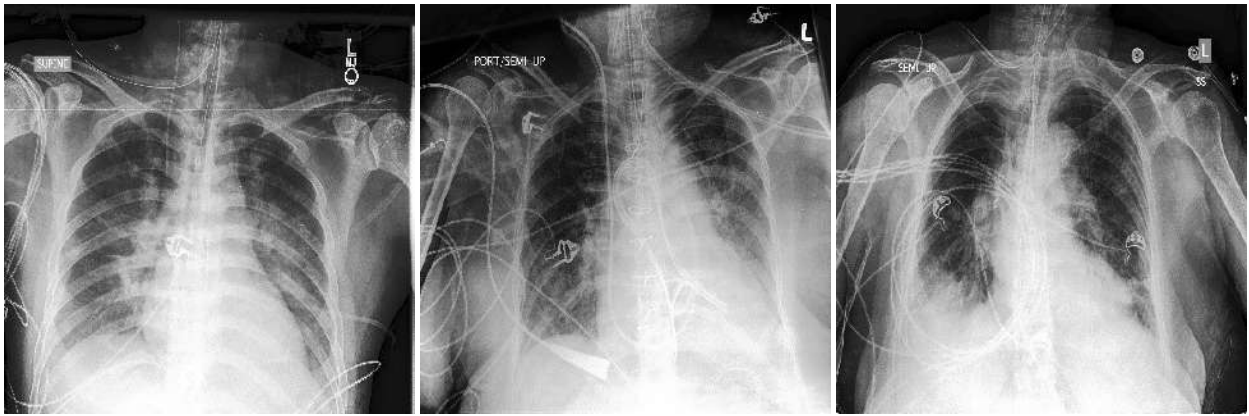

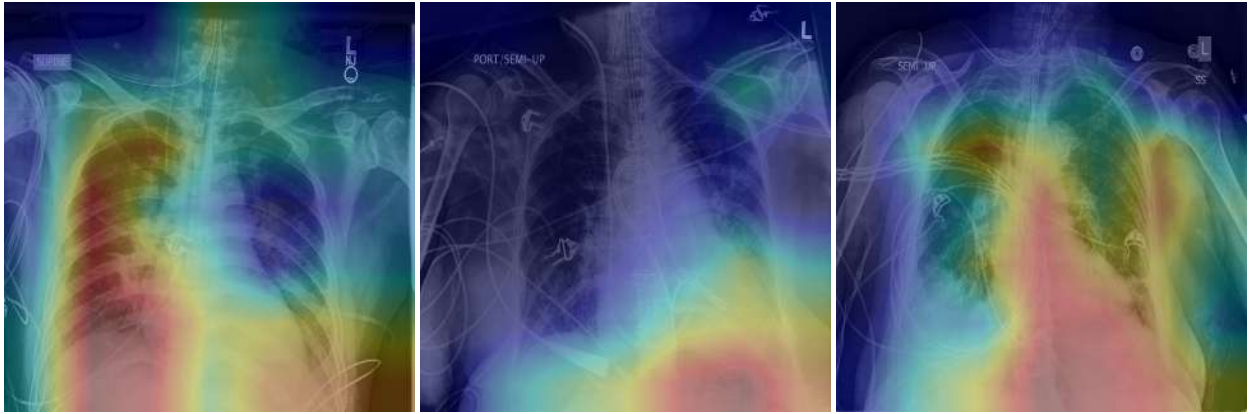

**Figure 4:** GRAD-CAM images of false positive cases detected by the COVID-net model on the ChexPert dataset

The presence of support devices can significantly bias models for many reasons: 1) these devices are much more visually apparent than infectious pulmonary findings (which is the true abnormality of interest), 2) they are seen in predictable anatomic locations, and 3) they are seen in predictable subsets of patients (i.e. ICU patients). If not accounted for in the model results, this would represent a significant source of bias as the models may learn a correlation between support devices and underlying infection rather than identification of the infection itself. Both the ChestX-ray14 and EMORY datasets did not contain annotations to identify the presence of support devices.

With respect to the high FN rate on the EMORY dataset (39 % for COVID-Net and 94 % for CoroNet), at least some of these errors could be attributed to the fact that no imaging findings are present despite a positive RT-PCR results. Based on the authors' clinical experience, there are a subset of patients which test positive for COVID-19 on serology but do not have appreciable findings on CXR. We are investigating this further in ongoing work to determine the rate of negative CXRs for patients with positive RT-PCR for COVID-19. This is because the ground truth on the EMORY dataset is based on RT-PCR results, rather than radiographic findings. Based on the authors' experience, there are a subset of patients with COVID-19 that either do not demonstrate findings on CXR, or demonstrate findings that are non-specific. Our results highlight the challenge of utilizing CXR for screening purposes only where RT-PCR is unavailable.

In regards to adequate training data, there is presently a bottleneck of data available for training models to detect COVID-19 that can be attributed to the relatively recent onset of the disease, HIPAA constraints for sharing data, and limited collaborative effort. The problem of overfitting and poor generalization is not limited to these two models or to the diagnosis of COVID-19. Rather, our results highlight a much broader issue in radiology AI where algorithms are frequently trained on narrow datasets and their performance suffers on outside data. Diverse selection of training data should include multiple sites, disease types, and careful consideration of the population to be evaluated (i.e. ED patients vs ICU patients) to maximize the potential for sustained model performance upon deployment.

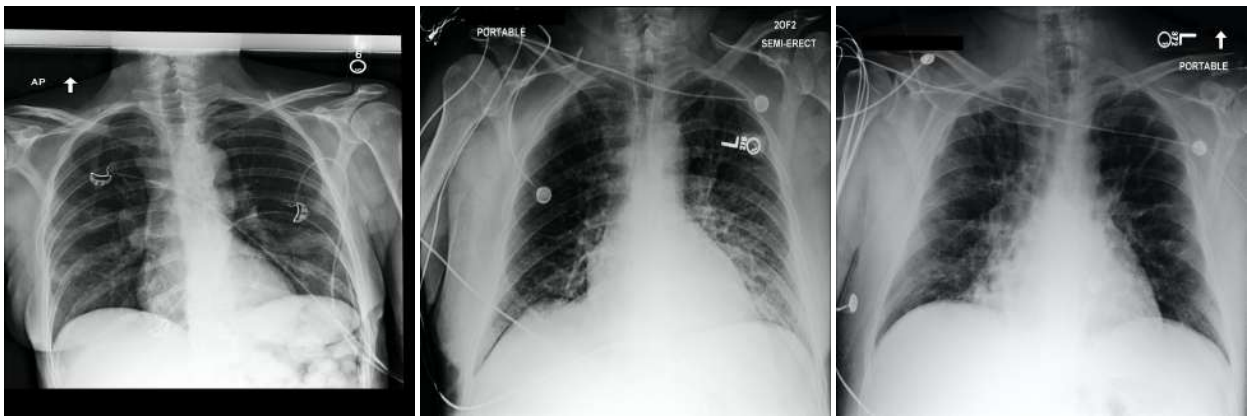

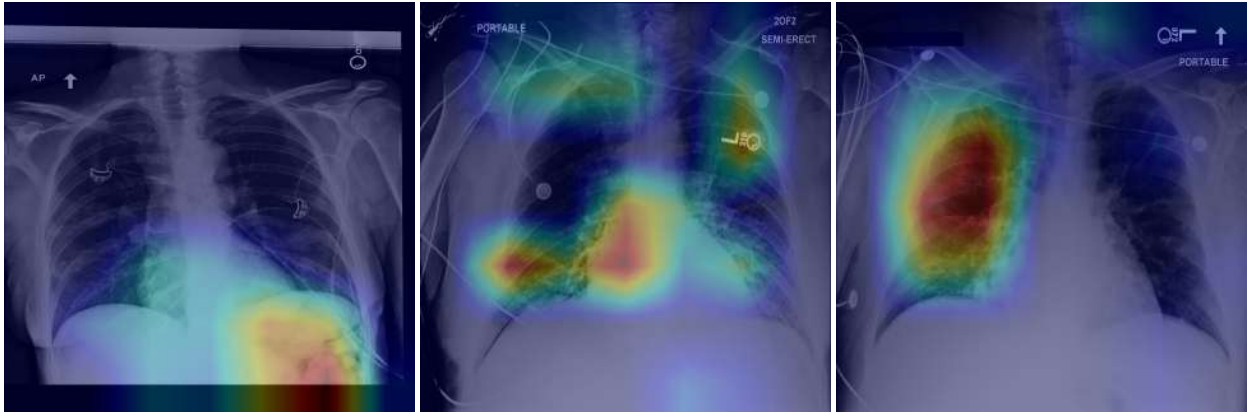

**Figure 5:** GRAD-CAM images of false positive cases detected by the COVID-net model in the MIMIC-CXR dataset

Lastly, COVID-19 is not an imaging diagnosis alone, but rather requires incorporation of multiple elements including physical exam, laboratory values, comorbidities, and clinical suspicion of disease. In order to build appropriate DL models to detect COVID-19, these additional elements should be included. We believe that a fusion model that incorporates these data elements from the EHR will offer better performance. Adoption of COVID-Net and CoroNet into clinical workflow as is would place a high burden on the radiologist who has to verify all cases due to high positive rates, and capture the missing cases in case of high false negative rate. We therefore advocate to develop AI tools that work together in partnership with humans to provide better performance. We also provide evidence that reporting expert level or better than expert performance does not equate to clinical utility and only perpetuates the hype of AI in radiology[20].

*Limitations*

Our study has several limitations. Firstly, the mechanism of ground labels differ across the four datasets used. We overcome this limitation by focusing on the task of COVID-19 diagnosis, which would be absent in the datasets published before December, 2019 when the first case was reported. There is debate on the role of imaging for COVID19 screening, hence we used the RT-PCR labeling as a proxy for ground truth labels on the EMORY dataset. Our second limitation is with regards to the explainability of the model inference using GRAD-CAM for visualization. Saliency maps show areas that contribute the highest probability to the model, and have been shown to be inaccurate in some cases. Our focus on this paper was not to explore the performance of various CNN interpretability models, and this can be an area of future research.

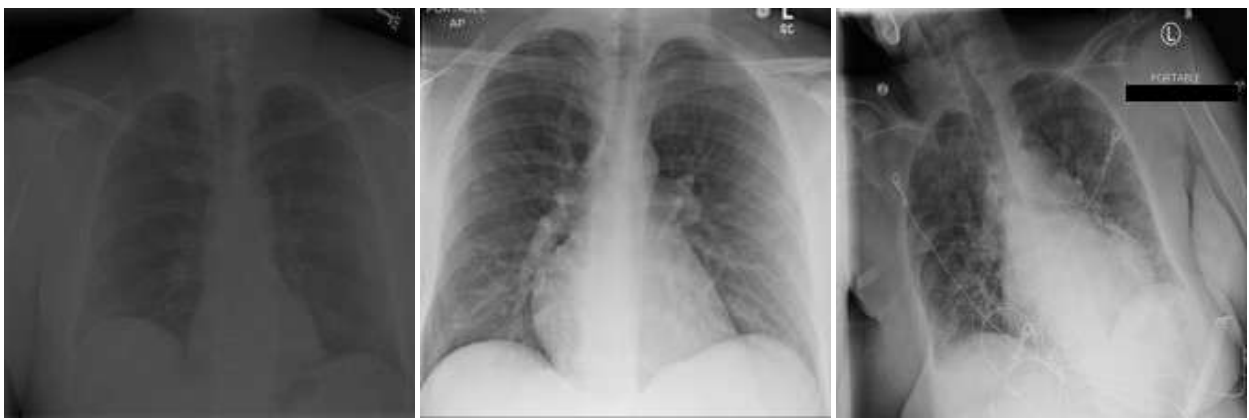

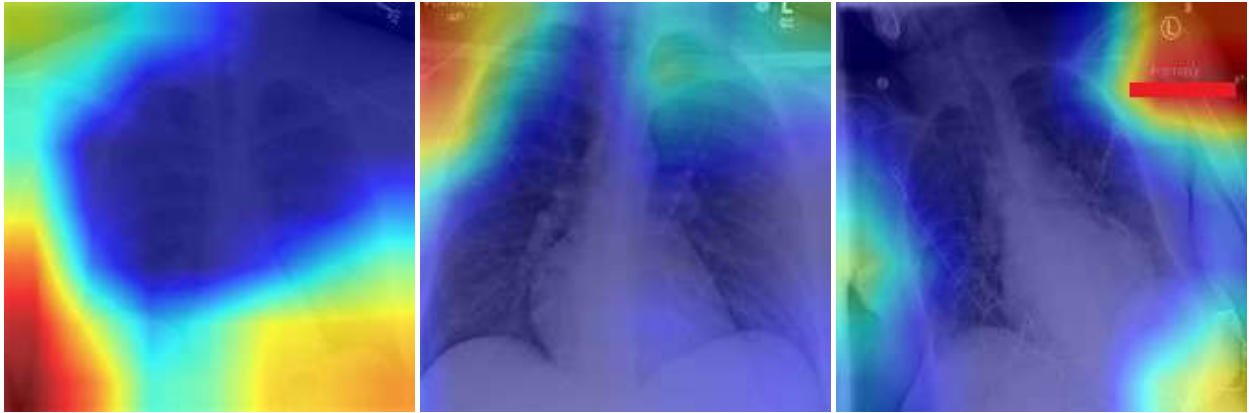

**Figure 6:** GRAD-CAM images of false negative cases predicted by the CoroNet model COVID-19 in the Emory dataset

## References

[1] *download*. https://www.fda.gov/media/134922/download. (Accessed on 06/22/2020).

[2] JCL Rodrigues et al. "An update on COVID-19 for the radiologist-A British society of Thoracic Imaging statement". In: *Clinical radiology* 75.5 (2020), pp. 323–325.

[3] Geoffrey D Rubin et al. "The role of chest imaging in patient management during the COVID-19 pandemic: a multinational consensus statement from the Fleischner Society". In: *Chest* (2020).

[4] Scott Simpson et al. "Radiological Society of North America Expert Consensus Statement on Reporting Chest CT Findings Related to COVID-19. Endorsed by the Society of Thoracic Radiology, the American College of Radiology, and RSNA." In: *Radiology: Cardiothoracic Imaging* 2.2 (2020), e200152.

[5] Pranav Rajpurkar et al. "Chexnet: Radiologist-level pneumonia detection on chest x-rays with deep learning". In: *arXiv preprint arXiv:1711.05225* (2017).

[6] Anna Majkowska et al. "Chest radiograph interpretation with deep learning models: assessment with radiologistadjudicated reference standards and population-adjusted evaluation". In: *Radiology* 294.2 (2020), pp. 421–431.

[7] Shinjini Kundu, Hesham Elhalawani, Judy W Gichoya, and Charles E Kahn Jr. *How Might AI and Chest Imaging Help Unravel COVID-19's Mysteries?* 2020.

[8] Alistair Mackenzie et al. "Breast cancer detection rates using four different types of mammography detectors". In: *European radiology* 26.3 (2016), pp. 874–883.

[9] Robert Koprowski. "Quantitative assessment of the impact of biomedical image acquisition on the results obtained from image analysis and processing". In: *Biomedical engineering online* 13.1 (2014), p. 93.

[10] Lucy M Warren et al. "Effect of image quality on calcification detection in digital mammography". In: *Medical physics* 39.6Part1 (2012), pp. 3202–3213.

[11] Carl F Sabottke and Bradley M Spieler. "The effect of image resolution on deep learning in radiography". In: *Radiology: Artificial Intelligence* 2.1 (2020), e190015.

[12] John R Zech, Marcus A Badgeley, Manway Liu, Anthony B Costa, Joseph J Titano, and Eric Karl Oermann. "Variable generalization performance of a deep learning model to detect pneumonia in chest radiographs: a cross-sectional study". In: *PLoS medicine* 15.11 (2018).

[13] Linda Wang and Alexander Wong. "COVID-Net: A Tailored Deep Convolutional Neural Network Design for Detection of COVID-19 Cases from Chest X-Ray Images". In: *arXiv preprint arXiv:2003.09871* (2020).

[14] Shahin Khobahi, Chirag Agarwal, and Mojtaba Soltanalian. "CoroNet: A Deep Network Architecture for SemiSupervised Task-Based Identification of COVID-19 from Chest X-ray Images". In: *medRxiv* (2020).

[15] Joseph Paul Cohen, Paul Morrison, and Lan Dao. "COVID-19 image data collection". In: *arXiv 2003.11597* (2020). URL: https://github.com/ieee8023/covid-chestxray-dataset.

[16] Yan-Rong Guo et al. "The origin, transmission and clinical therapies on coronavirus disease 2019 (COVID-19) outbreak–an update on the status". In: *Military Medical Research* 7.1 (2020), pp. 1–10.

[17] Ramprasaath R Selvaraju, Michael Cogswell, Abhishek Das, Ramakrishna Vedantam, Devi Parikh, and Dhruv Batra. "Grad-cam: Visual explanations from deep networks via gradient-based localization". In: *Proceedings of the IEEE international conference on computer vision*. 2017, pp. 618–626.

[18] Xiaosong Wang, Yifan Peng, Le Lu, Zhiyong Lu, Mohammadhadi Bagheri, and Ronald M Summers. "Chestxray8: Hospital-scale chest x-ray database and benchmarks on weakly-supervised classification and localization of common thorax diseases". In: *Proceedings of the IEEE conference on computer vision and pattern recognition*. 2017, pp. 2097–2106.

[19] Ian Pan, Saurabh Agarwal, and Derek Merck. "Generalizable inter-institutional classification of abnormal chest radiographs using efficient convolutional neural networks". In: *Journal of digital imaging* 32.5 (2019), pp. 888–896.

[20] Judy W Gichoya, Siddhartha Nuthakki, Pallavi G Maity, and Saptarshi Purkayastha. "Phronesis of AI in radiology: Superhuman meets natural stupidity". In: *arXiv preprint arXiv:1803.11244* (2018).